\documentclass{JHEP3}
\def\sq{\hbox {\rlap{$\sqcap$}$\sqcup$}}
\def\sq{\hbox {\rlap{$\sqcap$}$\sqcup$}}
\def\R{ {\rm R \kern -.31cm I \kern .15cm}}
\def\C{ {\rm C \kern -.15cm \vrule width.5pt \kern .12cm}}
\def\Z{ {\rm Z \kern -.27cm \angle \kern .02cm}}
\def\N{ {\rm N \kern -.26cm \vrule width.4pt \kern .10cm}}
\def\1{{\rm 1\mskip-4.5mu l} }
\def\lsim{\raise0.3ex\hbox{$<$\kern-0.75em\raise-1.1ex\hbox{$\sim$}}}
\def\gsim{\raise0.3ex\hbox{$>$\kern-0.75em\raise-1.1ex\hbox{$\sim$}}}
\def\noi{\noindent}
\def\beq{\begin{equation}}   \def\eeq{\end{equation}}
\def\bea{\begin{eqnarray}}  \def\eea{\end{eqnarray}}
\def\nn{\nonumber}
\def\noi{\noindent}
\def\beeq{\begin{eqnarray}} \def\eeeq{\end{eqnarray}}

\renewcommand{\theequation}{\thesection.\arabic{equation}}
\newcounter{hran} \renewcommand{\thehran}{\thesection.\arabic{hran}}

\def\bmini{\setcounter{hran}{\value{equation}}
   \refstepcounter{hran}\setcounter{equation}{0}
   \renewcommand{\theequation}{\thehran\alph{equation}}\begin{eqnarray}}

\def\bminiG#1{\setcounter{hran}{\value{equation}}
\refstepcounter{hran}\setcounter{equation}{-1}
\renewcommand{\theequation}{\thehran\alph{equation}}
\refstepcounter{equation}\label{#1}\begin{eqnarray}}

\def\emini{\end{eqnarray}\relax\setcounter{equation}{\value{hran}}\renewcommand{\theequation}{\thesection.\arabic{equation}}}

\title{Radion and moduli stabilization from induced brane actions in
higher-dimensional brane worlds}
\author{Christos Charmousis and Ulrich Ellwanger\\ LPT, 
Universit\'e de Paris-Sud, B\^at. 210, 91405 Orsay CEDEX, France \\
E-mail: \email{Christos.Charmousis@th.u-psud.fr} and 
\email{Ulrich.Ellwanger@th.u-psud.fr}}

\abstract{We consider a $4+N$-dimensional brane world  with 2
co-dimension 1 branes in an empty bulk. The two branes have $N - 1$ of
their extra dimensions compactified on a sphere $S^{(N-1)}$, whereas
the ordinary  $4$ spacetime  directions are Poincar\'e invariant. An
essential input are induced stress-energy tensors on the branes
providing different tensions for the spherical and flat part of the
branes. The junction conditions -- notably through their extra
dimensional components --  fix both the distance between the branes as
well as the size of the sphere. As a result, we demonstrate, that there
are no scalar  Kaluza-Klein states at all (massless or massive), that
would correspond to a radion or a modulus field of $S^{(N-1)}$. We also
discuss the effect of induced Einstein terms on the branes and show
that their coefficients are bounded from above, otherwise they lead to
a graviton ghost.}

\preprint{LPT-Orsay-0360}

\begin{document}

\section{Introduction}

The braneworld idea -- where our 4 dimensional universe is part of a
higher dimensional spacetime -- is largely motivated from string theory
and has attracted a lot of attention in recent years (see \cite{early}
for early work). In most proposed  models standard matter particles are
confined on a $d=4$ dimensional braneworld whereas gravity propagates
in the whole of spacetime.  

Models with $N$ compactified extra dimensions often suffer, however,
from phenomenologically unacceptable massless scalars (moduli)
associated to fluctuations of the size and the shape of the
compactified manifold. 
For example, in a simple setup of a 5 dimensional spacetime with two
3-branes \cite{1r,2r} (and geometry $R^{(3,1)} \times
S^1/Z_2$) a massless scalar is present, which is related to
fluctuations of the distance between the branes \cite{3r,4r} and
denoted the radion. Additional extra dimensions, parallel to the branes
and compactified on a manifold $M^{(N-1)}$, lead then generically to
additional massless moduli.
 
Such models with two co-dimension 1 branes are motivated by the strong
coupling limit of the heterotic string \cite{5r,6r} and were formulated
originally in $N = 7$ extra dimensions (with 6 among them along two
9-branes). Once one compactifies $N - 1$ extra dimensions on
$M^{(N-1)}$, the resulting geometry is thus $R^{(3,1)} \times
M^{(N-1)} \times S^1/Z_2$. Even without including the additional
degrees of freedom from tensor fields (or their scalar duals) in the
bulk \cite{7r}, one expects such models to suffer both from massless
moduli and a massless radion. 

Already long ago it was proposed to stabilize moduli through
matter-induced quantum effects \cite{8r}, and proposals to stabilize
the radion range from classical matter in the bulk [10-12] to, again,
Casimir-like forces involving matter in the bulk \cite{12r} or on the
branes \cite{13r}. For recent proposals on stabilisation, based on bulk
scalar fields, see  \cite{fl}. 

In the present paper we consider pure gravity in a $4+N$-dimensional
empty bulk with two co-dimension 1 branes. The branes consist of a
$4$ dimensional Poincar\'e invariant part and of $N-1$ extra dimensions
compactified on a sphere (where $N>2$).  Thus the topology of the setup
is $R^{(3,1)} \times S^{(N-1)} \times S^1/Z_2$. 

The essential ingredient in our setup is the addition of terms on the
brane actions originating from quantum effects of matter living on the
branes. These terms will be sourcing the Einstein equations via the 
junction conditions warping the background spacetime. 
We consider in particular induced stress-energy tensors on the branes,
that are constant and diagonal. At first sight these ressemble the
"standard" brane tension (or cosmological constant). However, an
overall cosmological constant or tension  $\Lambda$ corresponds to a
stress-energy tensor $T_A^B = \delta_A^B \Lambda$, with indices $A$,
$B$ both in $R^{(3,1)}$ along the brane (denoted by $\{A, B\} = \{\mu ,
\nu \}$ subsequently) and in $S^{(N-1)}$ along the brane (denoted by
$\{A, B\} = \{\alpha , \beta \}$ subsequently). The components of an
induced stress-energy tensor \cite{13r} will, however, generically be
different along $R^{(3,1)}$ or $S^{(N-1)}$, respectively\footnote{We
are grateful to V. Rubakov for pointing this out.}, 
\bea \label{1.1e}
&&T^{(ind)\ \nu}_{\phantom{(ind)}\mu} = \delta_{\mu}^{\ \nu} \
\Lambda^{(ind,4)} \ , \nn \\
&&T^{(ind)\ \beta}_{\phantom{(ind)}\alpha} = \delta_{\alpha}^{\ \beta}
\ \Lambda^{(ind,S)} \ ,
\eea
with
\beq \label{1.2e}
\Lambda^{(ind, 4)} \not= \Lambda^{(ind,S)} \ .
\eeq
The origin of this difference lies in the fact, e.g.,  that expectation
values of diagonal elements of operators like $<\partial_{\mu} \varphi
\partial^{\nu}\varphi>$ or  $<\partial_{\alpha} \varphi
\partial^{\beta}\varphi>$ are generically different due to the
different physical modes on $R^{(3,1)}$ or $S^{(N-1)}$ \cite{13r}.
Alternatively, vevs of higher rank tensor fields could induce
stress-energy tensors satisfying eq. (\ref{1.2e}) at the classical
level. Due to (\ref{1.1e}) with (\ref{1.2e}) the induced
stress-energy tensors cannot be absorbed completely by a
redefinition of the brane tension $\Lambda$. \par

The aim of the present paper is to investigate, whether the inclusion
of the above induced terms and the subsequent spacetime they induce,
allows for a stabilised 3-brane universe. To this end we ask whether
both the Einstein equations in the empty bulk and the junction
conditions on the branes can be satisfied. We then study spin 2
fluctuations (following essentially the work of \cite{roberto}) and 
study the scalar fluctuations of this system.

The essential results are as follows: First, one fine tuning condition
among the tensions and stress-energy tensors of the branes has to be
satisfied. This corresponds to the usual fine tuning of the effective
4d cosmological constant as for example in the Randall-Sundrum model.
An essential difference here is, however, that only one fine tuning is
required rather than two (one in each brane).  Once this fine tuning is
carried out, both the distance between the branes and the size of
$S^{(N-1)}$  are fixed by the junction conditions. Hence, there are no
massless modes associated to the radion or the overall modulus of
$S^{(N-1)}$. In order for this mechanism to work the presence of
induced stress-energy tensors is crucial; if they are switched off, the
branes collapse and the 4 + N dimensional world volume vanishes.

Given the absence of massless scalars, it still could be that there are
modes with negative mass$^2$, i.e. that the configuration is unstable.
We show, however, that there are no 4d scalar modes at all 
-- that are zero modes on $S^{(N-1)}$ and correspond to the radion
and/or the modulus of $S^{(N-1)}$ -- which could have potentially
negative masses$^2$. Thus, the inclusion of induced stress-energy
tensors satisfying (\ref{1.2e}) on the branes represents an economic
and efficient mechanism in order to stabilize both the radion and
moduli in higher dimensional brane worlds.

In addition we study the effect of induced Einstein terms as considered
previously in \cite{14r,15r} (see also \cite{ig} for early work on
induced gravity from quantum matter effects).  Such curvature terms can
be used to generate 4d gravity even in the case of one brane only
embedded in an infinite volume flat bulk. They can lead to modification
of gravity at very large scales \cite{15r}, \cite{Deffayet} which is a
potentially interesting explanation to the dark energy component put
forward by cosmological observations. This mechanism requires, however,
a huge coefficient multiplying the induced Einstein term (compared to
the fundamental gravitational scale in the bulk) which yields in turn a
strong coupling problem \cite{Luty} (see also \cite{sc} for recent
developments).  We will see, however, that in the present setup these
coefficients have to be relatively small in order not to turn the 4d
graviton into a ghost. Here 4d gravity will follow simply from the
compactness of all extra dimensions.

The paper is organized as follows: In the next section we give the
action, and the gravitational background that solves both the Einstein
equations in the bulk (using previous results in \cite{16r},
\cite{roberto}) and the junction conditions on the branes (studied
previously in \cite{17r}). We also study the ``close brane limit'', ie.
relatively small induced terms on the branes, leading to a relatively
small distance between the branes or weak warping of spacetime. In this
limit it is simpler to study the constraints on the parameters of the
model. In section 3 we show that 4d gravity emerges and we  give the
expression for the 4d gravitational constant (the Planck scale) in
terms of the fundamental  parameters. Section 4 is devoted to a
stability analysis of the setup, based on the check for the absence of
tachyons (modes with  negative mass$^2$), with the result announced
above: There are no such modes at all. Details of the calculation
(notably the wave equations for the gravitational fluctuations in the
bulk, and their junction conditions) are referred to appendix A.  In
section 5 we study the effect of induced Einstein terms on the branes
and show that they cannot play a dominant role. In section 6 we
conclude with a summary and an outlook.

\section{Higher-dimensional brane worlds with induced stress-energy
tensors}

We consider a $D = 4 + N$ dimensional spacetime,  with $N - 1$
dimensions  compactified on a sphere $S^{(N-1)}$. Along 4 dimensions
spacetime is Poincar\'e invariant, whereas  1 dimension (perpendicular
to the $D-2$-branes) is modded out by a $Z_2$ symmetry. The resulting 
topology is
\beq
\label{2.1e}
R^{(3,1)} \times S^{(N-1)} \times S^1/Z_2 \ .
\eeq
Our conventions for the coordinates are as follows: $x^{\mu}$, $\mu  =
0 \dots 3$, are the 4 coordinates on $R^{(3,1)}$, $z^{\alpha}$, $\alpha
= 1 \dots N-1$,  are the angles on $S^{(N-1)}$, $y$ is the coordinate
on $S^1/Z_2$. Upper case  latin letters run over all $D = 4 + N$
coordinates: $A,  B, \dots  = \{\mu , \alpha , y \}$. We will consider 
2 timelike branes of dimension $4+(N-1)$ situated at $y_1 = 0$ and 
$y_2 = \widehat{y}$, respectively. Coordinates along the branes are
thus $\{x^{\mu},z^{\alpha}\}$.

The bulk action includes the usual Einstein term $R^{(D)}$, but {\it 
no} cosmological constant. The actions on the branes 1 and 2 include
the cosmological constants  $\Lambda_1$ and $\Lambda_2$, and terms that
give rise, in the Einstein equations, to induced stress-energy tensors
satisfying eqs. (\ref{1.1e}) and (\ref{1.2e}). They follow from the
coupling of  gravity to composite operators of matter under the matter
path integral. For simplicity we will not specify these operators, but
denote them simply by $\Gamma_{i}^{(ind)}$ for $i=1,2$, which are
defined by the generation of the induced stress-energy tensors
$\Lambda_i^{(ind,4)}$ (along $R^{(3,1)}$), $\Lambda_i^{(ind,S)}$ (along
$S^{(N-1)}$) in the Einstein equations below. Thus the action reads

\beq \label{2.2e}
S = \int d^4x\ d^{N-1}zdy \left \{ \sqrt{-g^{(D)}} \ {1 \over 2 
\kappa_D} \ R^{(D)} 
+ \sum_{i=1,2} \sqrt{-g^{(D-1)}} \ \delta (y - y_i) \left [ 
\Lambda_i + \Gamma_{i}^{(ind)} \right ] \right \} \ ,
\eeq

\noi and the relevant components of the Einstein equations are

\bminiG{2.3e}
\label{2.3ae}
{1 \over \kappa_D} G_{\mu\nu}^{(D)} = \sum_{i=1,2} \sqrt{{g^{(D-1)} 
\over g^{(D)}}} \delta (y-y_i)
\left [ g_{\mu\nu} \Lambda_i + g_{\mu\nu} \Lambda_i^{(ind,4)}
\right ] \ ,  
\eeeq
\beeq
\label{2.3be}
{1 \over \kappa_D} G_{\alpha\beta}^{(D)} = \sum_{i=1,2} 
\sqrt{{g^{(D-1)} \over g^{(D)}}} \delta (y-y_i)
\left [ g_{\alpha\beta} \Lambda_i + g_{\alpha\beta}
\Lambda_i^{(ind,S)} \right ] \ ,
\eeeq
\beeq
\label{2.3ce}
{1 \over \kappa_D} G_{yy}^{(D)} = {1 \over \kappa_D} G_{\mu y}^{(D)} 
= {1 \over \kappa_D}
G_{\alpha y}^{(D)} = 0
\emini

Let us first concentrate on the metric in the bulk, where the
right-hand  sides of eqs. (\ref{2.3e}) vanish. Given the spacetime
symmetries of our configuration the metric reads,
\beq
\label{2.4e}
ds^2 = A^2 (y) \eta_{\mu\nu} dx^{\mu} dx^{\nu} + B^2(y) d^2y + C^2(y) 
\gamma_{\alpha \beta}
dz^{\alpha} dz^{\beta} \eeq
and $\gamma_{\alpha\beta}(z)$ is the metric on $S^{(N-1)}$ with  radius
1.  Solutions of the Einstein equations in the bulk for a metric of
the  form (\ref{2.4e}) have been obtained in \cite{16r}. Our form of
the solutions is  related to the ones in \cite{16r} (see also
\cite{roberto}) by a simple redefinition of the ``radial'' coordinate
$y$, such that --  in the present case -- the action of the $Z_2$
symmetry becomes simply $y \to - y$.  Defining a $Z_2$ invariant
function $f(|y|)$,
\beq \label{2.5e}
f(|y|) = 1 + \alpha\ e^{-(N-2) {|y| \over r_0}} 
\eeq
the solution reads,
\bea \label{2.6e}
&&A(y) = f(|y|)^{n_A}\quad , \quad n_A = - {1 \over 4} \sqrt{{N-1 
\over N+2}} \ , \nn \\
&&B(y) = e^{{|y| \over r_0}}\ f(|y|)^{n_B}\quad , \quad n_B =  {1 
\over N-2} \left ( \sqrt{{N-1 \over
N+2}} - {N-3 \over 2} \right )  \ , \nn \\
&&C(y) = r_0\ e^{{|y| \over r_0}}\ f(|y|)^{n_C}\quad , \quad n_C = 
{1 \over N-2} \left ( \sqrt{{N-1
\over N+2}} + {1 \over 2} \right )  \label{2.3f}\ .
\eea

A few comments are now in order. Clearly the parameter $\alpha$ in eq.
(\ref{2.5e}) can be gauged to 1 in the bulk,  since one could 
eliminate it by a shift in $y$. However, the addition of brane 1 in the
"gauge" $y_1 = 0$ breaks translational invariance in $y$, and then
$\alpha$ has to be kept as a parameter. 

A possible solution in a bulk without branes (before modding out by
$Z_2$) is given by $f(y)=1 + \alpha e^{-(N-2) {y \over r_0}}$ and is
therefore singular for $y\rightarrow -\infty$. We are cutting off this
part of spacetime by introducing a brane at $y=y_1=0$. Accordingly
parameter $\alpha$ dictates the portion of the bulk spacetime we cut
off. A second parameter $r_0>0$  is related to the ADM energy of the
background metric (\ref{2.4e}) much like a mass parameter in a black
hole solution. It also characterises the size of the spheres, cf. the
expression for $C(y)$ in (\ref{2.3f}). Finally, a third parameter
consists in the position $y_2 = \widehat{y}$ of the second brane.

The solutions are well defined only for $N>2$ since it is only then that
the $(N-1)$-spherical subspace has nontrivial 
curvature. Finally the bulk metric would
be asymptotically flat as $y\rightarrow \infty$. To put it in a
nutshell  the addition of the branes at $y_1=0$ and $y_2=\widehat{y}$ 
cuts off the singular part of spacetime and asymptotic infinity
respectively. Therefore the coordinate $y$ varies over the finite
interval  $0 \leq |y| \leq \widehat{y}$ where the bulk metric is always
non-singular. Finally $Z_2$ symmetry requires all functions $A(y)$,
$B(y)$ and $C(y)$ to be even under $y \to - y$.

Next we have to consider the junction conditions, which follow from 
matching the singular terms in eqs. (\ref{2.3ae}) (along $R^{(3,1)}$)
and (\ref{2.3be}) (along $S^{(N-1)}$) for $y \to y_i$. At each brane
one obtains two independent equations corresponding  to eqs.
(\ref{2.3ae}) and (\ref{2.3be}), respectively. They read,
\bminiG{2.7e}
\label{2.7ae}
{1 \over \kappa_D B_i} \left ( - 3 \left [ {A'' \over A}\right ]_i - 
(N-1) \left [ {C'' \over
C}\right ]_i \right ) =  \Lambda_i + \Lambda_i^{(ind, 4)} \ ,
\eeeq
\beeq
  \label{2.7be}
{1 \over \kappa_D B_i} \left ( - 4 \left [ {A'' \over A}\right ]_i - 
(N-2) \left [ {C'' \over
C}\right ]_i \right ) = \Lambda_i + \Lambda_i^{(ind,S)}  ,
\emini
where
\beq \label{2.8e}
\left [ {A'' \over A} \right ]_i = \left ( {\partial_y A \over A} 
\right |_{y_i+ \varepsilon} \left. - \
{\partial_y A \over A} \right |_{y_i - \varepsilon} \left. \right ) = 2 
\sigma_i {A'_i \over A_i} \eeq
\noi and similarly for $[C''/C]$. The last equality in  (\ref{2.8e})
holds due to the $Z_2$ symmetry, and the sign $\sigma_i$ is $\sigma_1 =
+ 1$, $\sigma_2 = -  1$. $B_i$ denotes simply $B(y_i)$ etc.  Using the
expressions (\ref{2.6e}) for $A(y)$, $B(y)$ and $C(y)$,  equations
(\ref{2.7e}) can be rewritten as
\bminiG{2.9e}
\label{2.9ae}
\sqrt{f_i} (1 - w) + {1 \over \sqrt{f_i}} (1 + w) =
\sigma_i \kappa_D C_i \left ( \Lambda_i^{(ind,S)} -
\Lambda_i^{(ind,4)}\right ) \ , \eeeq
\beeq
\label{2.9be}
&\sqrt{f_i} &\left ( N-1 + (N-3)w \right ) + {1 \over \sqrt{f_i}} 
\left ( N-1 - (N-3)w \right )  \nn
\\ &&= -
\sigma_i \kappa_D C_i \left ( 2 \Lambda_i + (N-1) 
\Lambda_i^{(ind,S)}- (N-3) \Lambda_i^{(ind, 4)}\right )
  \emini
\noi where $f_i = f(|y_i|)$, and
\beq \label{2.10e}
w = {1 \over 2} \sqrt{(N-1) (N+2)} \ .
\eeq

It is important to compare the number of equations to solve  with the
number of free parameters: Equations (\ref{2.9e}) have to be satisfied
at both  branes 1 and 2, which gives 4 equations altogether. On the
other hand the configuration has just 3 free  parameters: $r_0$,
$\alpha$ in the metric (in the function $f$ in (\ref{2.5e}) and
(\ref{2.6e})) and the position $\widehat{y}$ of the second  brane.
Hence one fine tuning condition among the bare parameters $\Lambda_i$,
$\Lambda_i^{(ind)}$, $i = 1, 2$ in (\ref{2.9e}), is required for a 
solution to exist. This is nothing but the ``usual'' fine tuning of the
cosmological constant. Here we  have assumed that the gravitational
background is Poincar\'e invariant along 3+1 dimensions;  if the fine
tuning condition was not satisfied, we would find 3+1 de-Sitter or
anti-de-Sitter solutions. 

Note, however, that here we have ``just'' {\it one} fine tuning to 
perform compared to {\it two} fine tunings that are required in the
Randall-Sundrum model \cite{1r}  with a cosmological constant in the
bulk. The need to perform two fine tunings in this latter  model can be
traced back to the particular form of the logarithmic derivative of the
warp factor in a portion of AdS space, which is {\it independent} of
$y$. In this case the distance between the branes never enters the
junction conditions, and  consequently it is left free (corresponding
to the massless radion), but two fine  tunings are required. Here the
junction conditions {\it fix} $\widehat{y}$, the distance  between the
branes. Hence there is no massless radion. Moreover, the radius of the
compact space  $S^{(N-1)}$ is fixed: According to the metric
(\ref{2.3e}) it is given by $C(y)$ (and varies thus  with $y$), but
since all parameters as $r_0$ and $\alpha$ are determined by eqs.
(\ref{2.9e}), $C(y)$ is completely fixed. Hence there is no massless
modulus related to a fluctuating size of  $S^{(N-1)}$. 

The explicit expressions for the bulk parameters $r_0$, $\alpha$ and 
$\widehat{y}$ as functions of the bare parameters on the branes 
can be obtained from
(\ref{2.9e}). Defining,
\beq
\label{2.11e}
L_i \equiv 2 \Lambda_i + (N-1) \Lambda_i^{(ind,S)} - (N-3)
\Lambda_i^{(ind, 4)}, \qquad  
\delta_i \equiv  {\Lambda_i^{(ind, S)} - \Lambda_i^{(ind,  4)} 
\over L_i} 
\eeq
and
\beq
\label{2.12e}
v = N-1 + (N-3)w
\eeq
(with $w$ as in eq. (\ref{2.10e})) we obtain
\beq\label{2.13e} 
\alpha={2(1+(N-1)\delta_1)\over w-1-v\delta_1} 
\eeq
\beq\label{2.14e} 
r_0=-{2(N-1)+v\alpha\over \kappa_D (1+\alpha)^{n_c+{1\over 2}} L_1}
\eeq
with $n_C$ as in eqs. (2.6). In addition it is useful to define
\bea 
s_1 = {1+(N-1)\delta_1 \over 1+(N-1)\delta_2},\quad
s_2 = {w-1-v\delta_1 \over w-1-v\delta_2}, \quad
\label{2.15e}
s_3 = {1+w+(2(N-1)-v)\delta_1 \over 1+w+(2(N-1)-v)\delta_2},
\eea
which gives
\beq\label{2.16e}
e^{(N-2){|\widehat{y}|\over r_0}}= {s_1 \over s_2}
\eeq
and the fine tuning condition
\beq\label{2.17e}
{L_1 \over L_2} = - {1 \over s_2} \left({s_1 \over s_2}\right)^{1 \over
N-2}  \left({s_2 \over s_3}\right)^{n_c + {1 \over 2}}
\eeq

Note that if the induced terms are equal ($\delta_i \to 0$ for $i =
1,2$) eq. (\ref{2.15e}) leads to $s_1 = s_2 = s_3 = 1$, and thus eq.
(\ref{2.16e}) implies $|\widehat{y}|=0$. Hence, without the differing 
induced terms, the distance between the branes vanishes and the branes
collapse. Non-vanishing  induced terms -- at least on one brane -- are
necessary for the present setup to exist. 

Let us now consider the approximation $\Lambda_i^{ind}/\Lambda_i \ll
1$, i.e. relatively small induced stress-energy tensors, which
simplifies the above equations. From eq. (\ref{2.11e}) this implies 
\beq  \label{2.19e}
|\delta_i| \ll 1
\eeq
both for $i = 1$ and 2. Then the junction conditions (\ref{2.9e}) give
the following results to leading order in $\delta_i$: The fine tuning
condition (\ref{2.17e}) reads
\beq
\label{2.21e}
\Lambda_2 = - \Lambda_1 \ ,
\eeq
and for the parameters $\alpha$, $r_0$ and $\widehat{y}$ one  obtains
(with $w$ as in eq. (\ref{2.10e}))
\bminiG{2.22e}
\label{2.22ae}
\alpha = {2 \over w- 1} \ ,
\eeeq
\beeq
  \label{2.22be}
r_0 = - {2w(N-2) \over \kappa_D \Lambda_1 (w+ 1)} \ \left ( {w-1 
\over w+1}\right )^{n_C - {1 \over
2}} \ ,
  \eeeq
\beeq
  \label{2.22ce}
|\widehat{y}| = 
{2wr_0 \over w-1} \left ( \delta_1 - \delta_2 \right )\ .
\emini
In view of eq. (2.18) the distance between the branes is thus relatively
small, and we will denote this situation as the "close brane limit" (to
be studied again in section 5).
Clearly, $r_0 > 0$ implies, from eqs. (\ref{2.22be}) and
(\ref{2.21e}),  that $\Lambda_1 < 0$ and $\Lambda_2 > 0$, hence brane~1
is the negative tension brane. (This asymmetry originates from our
convention of a negative exponent in the function $f$ in eq.
(\ref{2.5e}).)  Here the leading order expressions for $\delta_i$ are
\beq
\label{2.25e}
\delta_i \cong {1 \over 2 \Lambda_i} \left ( \Lambda_i^{(ind,S)} - 
\Lambda_i^{(ind,4)} \right ) \ .
\eeq
Hence the positivity of the right-hand side of eq. (\ref{2.22ce}) 
implies (with $\Lambda_1 < 0$,
$\Lambda_2 > 0$)
\beq
\label{2.26e}
\Lambda_1^{(ind,4)} + \Lambda_2^{(ind,4)} > \Lambda_1^{(ind,S)} + 
\Lambda_2^{(ind,S)} \ .
\eeq
The inequality (\ref{2.26e}) is not difficult to fulfill, and is 
sufficient for the existence of a consistent solution of the junction
conditions. 

In the next section we consider the gravitational fluctuations around 
the background (\ref{2.4e}), and concentrate on the emergence of the 4d
graviton.

\section{Background fluctuations and 4d gravity}

In order to obtain the linear equations for gravitational fluctuations,
we expand the metric about the curved background $g_{AB}^{(0)}$ 
(\ref{2.4e}),
\beq \label{3.1e}
ds^2 = \left ( g_{AB}^{(0)} + h_{AB} \right ) dx^A dx^B \ ,
\eeq
and use the Einstein equations (\ref{2.3e}). D-dimensional general
coordinate invariance corresponds to the following gauge freedom:

\beq \label{3.2e}
h_{AB} \to h'_{AB} + \nabla_A \xi_B + \nabla_B \xi_A
\eeq
where $\nabla_A$ are covariant derivatives computed with
$g_{AB}^{(0)}$, and $\xi_A$ are $D = 4 + N$ arbitary functions. To fix
(partially) the  gauge freedom (\ref{3.2e}) we choose the axial gauge,

\beq \label{3.3e}
h_{\mu y} = h_{\alpha y} = 0 \ .
\eeq

It is easily checked that the junction conditions do not involve the
modes (\ref{3.3e}). Indeed the modes in (\ref{3.3e}) are odd under the
$Z_2$-symmetry, and  therefore vanish on both branes. In principle, the
freedom (\ref{3.2e}) also allows to gauge $h_{yy}$  to zero. However,
there are also (non-gravitational) modes, that correspond to
fluctuations of the brane positions around $y_1 = 0$, $y_2 = {\widehat
y}$. We choose to use the remaining gauge degree of freedom to gauge
these brane bending modes away, i.e. to keep the brane positions fixed.
Then $h_{yy}$ cannot be gauged away as well \cite{3r,4r}. 

With respect to a 4-dimensional observer bulk  metric fluctuations can
be seperated into scalar, spin 2 and ``vector'' like. Indeed the
latter   fluctuations with the index structure $h_{\mu\alpha}$
correspond to  4d vector fields. These will always contain massless
states in 4d, which form a Yang-Mills sector  with the gauge group
$O(N-1)$, the isometrics of $S^{(N-1)}$. Subsequently we will not be
interested in  this sector. We are left with fluctuations with the
index structures $h_{\mu\nu}$,  $h_{yy}$ and $h_{\alpha \beta}$, which
generally depend on $x^{\mu}$, $y$ and $z^{\alpha}$.  Furthermore,  we
confine ourselves to the lowest mass states with respect to the angular
mode excitations on $S^{(N-1)}$. The perturbations $h_{\mu\nu}$ and
$h_{yy}$ are then independent of the angles $z^{\alpha}$ on 
$S^{(N-1)}$, and $ h_{\alpha\beta}$ is proportional to the metric
$\gamma_{\alpha\beta}$ on $S^{(N-1)}$. It is convenient to rescale the
linear fluctuations $h_{A B}$  by the background metric, i.e. to define
$\widehat{h}_{\mu\nu}$, $\widehat{h}_S$, $\widehat{h}_{yy}$ as
\beq \label{3.4e}
h_{\mu\nu} (x^{\mu}, y) = A^2(y)\ \widehat{h}_{\mu\nu}(x^{\mu}, y) \ .
\eeq
\beq
\label{3.5e}
h_{\alpha\beta}(x^{\mu},y,z^{\alpha}) = {C^2(y) \over N-1} 
\gamma_{\alpha\beta}(z_{\alpha})
\ \widehat{h}_S(x^{\mu}, y) \eeq
\beq
\label{3.6e}
h_{yy}(x^{\mu}, y) = B^2(y) \ \widehat{h}_{yy} (x^{\mu}, y) \ .
\eeq
The scalar modes $\widehat{h}_S$ and $\widehat{h}_{yy}$ represent
fluctuations of the overall size of the sphere (the modulus) and the
radion mode respectively. 

The Einstein equations for the fluctuations can be split into  two
regimes: In the bulk, away from the branes, the vanishing of the $D$ 
dimensional Einstein tensor implies the vanishing of all components of
the Ricci tensor. In appendix A we give all components $R_{\mu\nu}$,
$R_{yy}$, $\gamma^{\alpha \beta} R_{\alpha \beta}$ and $R_{\mu y}$ of 
the Ricci tensor linear in the fluctuations $\widehat{h}_{\mu\nu}$,
$\widehat{h}_{yy}$ and  $\widehat{h}_S$. $R_{\mu y}$ is related to the
other components by a Bianchi identity, and  $R_{\alpha y}$ vanishes
identically for $z^{\alpha}$-independent fluctuations. On the branes we
have to consider the junction conditions for the  fluctuations which
are also given in appendix A. 

Let us now turn to the massless spin 2 fluctuations with respect to a
4-d observer. Our approach follows closely that of \cite{roberto}, the
essential difference here being the presence of two co-dimension 1
branes and of the induced curvature terms. These fluctuations contain a
massless transverse tracefree mode $h_{\mu\nu}^{(grav)}$ representing
the 4d graviton:

\beq \label{3.7e}
\partial^{\mu} h_{\mu\nu}^{(grav)} = \eta^{\mu\nu} 
h_{\mu\nu}^{(grav)} = 0 \ .
\eeq

The 4d graviton is easily identified \cite{roberto} as the $y$
independent mode of $\widehat{h}_{\mu\nu}$ satisfying (\ref{3.7e}).
Neglecting the scalar fluctuations ${\widehat h}_{\mu}^{\ \mu}$,
${\widehat h}_{yy}$ and ${\widehat h}_S$, the wave equations 
(A.5b)-(A.5d) in  appendix A are trivially satisfied. The wave equation
(A.5a), originating from the vanishing of $R_{\mu\nu}$, reduces to
\beq \label{3.8e}
\sq^{(4)} \ \widehat{h}_{\mu\nu} = 0 \ .
\eeq
The junction conditions (A.7) and (A.8) are also satisfied, and this
mode is trivially normalizable, since all extra dimensions are
compact. 

In order to obtain an expression for the 4d Planck mass we  proceed as
follows: First one defines a rescaled 4d metric $\widehat{g}_{\mu\nu}$
as
\beq \label{3.9e}
g_{\mu\nu} = A^2(y) \ \widehat{g}_{\mu\nu} \ .
\eeq
Then we insert the expression (\ref{3.9e}) for the metric into the 
action (\ref{2.2e}), and integrate over all extra dimensions $dy$,
$d^{N-1} z$.  The result is an effective 4d action of the form
\beq \label{3.10e}
S_{eff} = \int d^4x\ \sqrt{-\widehat{g}} \ {1 \over 2 \kappa_4} \ 
\widehat{R}^{(4)}
\eeq
where $\widehat{R}^{(4)}$ is the Ricci scalar constructed from 
$\widehat{g}_{\mu\nu}$. Comparing the coefficients, one obtains for
$\kappa_4$
\beq \label{3.11e}
{1 \over \kappa_4} = \Omega^{N-1} {2 \over \kappa_D} 
\int_{0}^{\widehat{y}} dy\ A^2(y) \ B(y) \ C(y)^{N-1}
\eeq
where $\Omega^{N-1}$ is the volume of $S^{(N-1)}$ with radius 1,  and
the factor 2 in front of the integral over $dy$ originates from the
fact that originally it ranges  from $-\widehat{y}$ to $+
\widehat{y}$. 
This integral is always finite provided $\widehat{y}$ is finite and
corresponds to the normalisation factor for the constant  4-dimensional
graviton mode (and the subsequent massive spin 2 modes). The analysis
of the massive spin 2 modes and higher angular momenta  goes through as
in \cite{roberto} modulo the presence of the boundary terms. The spin 2
modes are stable provided the right hand side of eq. (\ref{3.11e}) is
positive. 

This is not quite the end of the story. Assuming that matter lives on 
one of the two branes 1 or 2, the original coupling of matter to
gravity is of the form
\beq \label{3.12e}
\left . \int d^4x \ d^{N-1} z \sqrt{-g^{(D-1)}} \ {\cal L}_{matter} 
\left ( g^{(D-1)}\right ) \right
|_i \ .
  \eeq
Assuming that the matter wave functions are zero modes on  $S^{(N-1)}$,
the $d^{N-1}z$ integral can be performed as before which turns
(\ref{3.12e}) into
\beq \label{3.13e}
\Omega^{N-1}\ C_i^{N-1} \int d^4x \sqrt{-\widehat{g}} \ A_i^4\ {\cal 
L}_{matter} \left
( A_i^2 \widehat{g}_{\mu \nu}\right )  \ .
  \eeq
The powers of $A_i \equiv A(y_i)$ cancel exactly in (\ref{3.13e}) for 
scale invariant matter (kinetic terms of gauge fields), but have to be
eliminated by a field  redefinition in the case of non-scale invariant
matter as scalars. As proposed in \cite{1r} this could  generate mass
scales that are naturally much smaller than the fundamental
(gravitational) scales of  the theory, if $A_i$ is very small. We will
not consider this possibility here.  The prefactors $\Omega^{N-1}
C_i^{N-1}$ in (\ref{3.13e}) can also be  eliminated by a rescaling of
the fields (such that their kinetic terms are properly normalized)  and
the couplings in the action. Recall that with our convention
(\ref{2.4e}) for the metric, where  $z^{\alpha}$ are dimensionless
angles on $S^{(N-1)}$, $C(y)$ has the dimension of a length (cf. the 
last of eqs. (\ref{2.6e})). Only after this last rescaling the fields
and couplings assume their dimensions appropriate for a 4d effective
theory. 

\section{Stability analysis of the scalar fluctuations}

The absence of both a massless radion and a massless modulus associated
to the size of $S^{(N-1)}$ follows from the fact that both the distance
$\widehat{y}$ between the branes and the ($y$-dependent) radius $C(y)$
of $S^{(N-1)}$ are fixed, through the junction conditions, in terms of
the parameters in the action. However, these (or other) scalar modes
could acquire a negative (4d) mass$^2$, which would signal a tachyonic
instability of the present configuration.\par

In order to check this, we have to study the lowest 4d scalar
Kaluza-Klein states contained in the gravitational fluctuations
$h_{\mu\nu}$, $h_{yy}$ and $h_{\alpha\beta}$. As indicated in eqs.
(3.4)-(3.8) in the previous section, we confine ourselves to zero modes
on $S^{(N-1)}$. The crucial question is then whether there are modes
$\varphi (x^{\mu}, y)$ (Kaluza-Klein states along the $y$ direction)
with possibly negative 4d mass$^2$. Our important result is that,
somewhat astonishingly, there are no such modes at all (beyond pure
gauge transformations), hence the present configuration is extremely
rigid. This result is far from obvious, and will follow from a detailed
analysis of the bulk equations of motion and the junction conditions.
\par

First, the gravitational fluctuations $h_{\mu\nu}$, $h_{yy}$ and
$h_{\alpha\beta}$ do not represent the complete spectrum of potential
physical fluctuations of the configuration: In addition there are brane
bending modes $\beta_i (x^{\mu})$, which correspond to fluctuations 
$y_i \to y_i + \beta_i (x^{\mu})$ around the positions $y_i \equiv
\{y_1, y_2\} = \{0, \widehat{y}\}$ of the branes (in a given coordinate
system).\par

However, the gauge (3.3)

\beq \label{4.1e} h_{\mu y} = h_{\alpha y} = 0 \eeq

\noi still allows for general infinitesimal coordinate transformations

\beq \label{4.2e} \delta x^{\mu} = \xi^{\mu} (x^{\mu} , y) \ , \quad
\delta y = \chi (x^{\mu}, y) \eeq

\noi provided that

\beq \label{4.3e} \eta_{\mu \nu} \ {d\xi^{\nu} \over dy} = - {B^2 \over
A^2}\ {d\chi \over dx^{\mu}} \eeq

\noi relating $\xi^\mu$ and $\chi$ so  that the gauge condition
(\ref{4.1e}) remains valid.  This  gauge freedom (\ref{4.2e}) can be
partially used to gauge away the brane bending modes $\beta_i$, by a
suitable choice of $\chi (x^{\mu}, y_i)$. In fact we  are still left
with general coordinate transformations of the form (\ref{4.2e}) with
(\ref{4.3e}), as long as the gauge parameter $\chi (x^{\mu}, y)$ {\it
vanishes on both branes}, i.e. satisfies Dirichlet boundary conditions.
This boundary gauge choice is used implicitely in the junction
conditions (A.13)-(A.15) in the appendix, where the brane bending modes
$\beta_i$ have already been omitted.\par

It is convenient to decompose the gravitational fluctuations into plane
waves along $R^{(3,1)}$, i.e.

\beq \label{4.4e} h_{\mu\nu}, h_{yy}, h_{\alpha \beta} \sim
e^{ik_{\lambda}x^{\lambda}} \eeq

\noi where $k^2 = - m^2$, the mass of the mode in question (and one can
assume $m^2 \not= 0$). \par

Out of these fluctuations we can construct four 4d scalars (and zero
modes on $S^{(N-1)}$): $\widehat{h}_S$ and $\widehat{h}_{yy}$ have
already been defined in eqs. (3.5) and (3.6), and out of
$\widehat{h}_{\mu\nu} = A^{-2} h_{\mu\nu}$ we can project

\beq \label{4.5e} \widehat{h}_4 = \eta^{\mu\nu} \ \widehat{h}_{\mu\nu}
\eeq

\noi and, for non-zero modes,

\beq \label{4.6e} \widehat{L} = k^{-2} k^{\mu} k^{\nu}
\widehat{h}_{\mu\nu} = - m^{-2} k^{\mu} k^{\nu} 
\widehat{h}_{\mu\nu} \eeq

\noi (hence $L$, as defined in eq. (A.6), is related to $\widehat{L}$
through $L = m^2A^{-2} \widehat{L}$).

The remaining general coordinate transformations (\ref{4.2e}) with
(\ref{4.3e}) (and Dirichlet boundary conditions on $\chi$) act on the
four 4d scalars $\widehat{h}_4$, $\widehat{L}$, $\widehat{h}_{yy}$ and
$\widehat{h}_S$ with masses $m^2$ as

\bminiG{4.7e}
\label{4.7ae}
\delta \widehat{h}_4(y) = 2m^2 \int_0^y dy' \ {B^2 \over A^2} \ \chi 
- 8 {A' \over A}\ \chi \ ,
   \eeeq
\beeq
  \label{4.7be}
\delta \widehat{L}(y) = 2m^2 \int_0^y dy' \ {B^2 \over A^2} \ \chi - 2 
{A' \over A}\ \chi \ ,
  \eeeq
\beeq
  \label{4.7ce}
\delta \widehat{h}_{yy}(y) = -2\left ( \chi ' + {B' \over B}\ \chi
\right ) \  ,
\eeeq
\beeq
  \label{4.7de}
\delta \widehat{h}_{S}(y) = -2 ( N-1 ) {C' \over C}\ \chi \  .
\emini

\noi The bulk equations given in the appendix simplify somewhat when 
written in terms of the four modes above. First we define

\beq
\label{4.8e}
a = A^4\ C^{N-1} \ B^{-1}
\eeq

\noi such that

\beq
\label{4.9e}
D' \equiv 4 {A' \over A} + (N-1){C' \over C} - {B' \over B} = a'a^{-1} \ ,
\eeq

\noi and in addition we define

\beq
\label{4.10e}
E' = 3 {A' \over A} + (N-1) {C' \over C} \ .
\eeq

\noi Then, in an obvious notation, the bulk equations in the appendix
can be rewritten as follows:\par

\noi $2aB^2 \times$ (A.7) becomes

\beq \label{4.11e} [a\widehat{h}'_4 ]' + 4a {A' \over A} \left (
\widehat{h}'_4 + \widehat{h}'_S - \widehat{h}'_{yy} \right ) + a {B^2
\over A^2} m^2 \left ( 2 \widehat{h}_4 - 2 \widehat{L} + \widehat{h}_S
+ \widehat{h}_{yy}\right ) = 0 \ , \eeq

\noi $2aB^2 {A^2 \over m^2} \times$ (A.8) becomes

\beq
\label{4.12e}
[a\widehat{L}' ]' + a {A' \over A} \left (
\widehat{h}'_4 + \widehat{h}'_S - \widehat{h}'_{yy} \right ) + a {B^2
\over A^2} m^2 \left (  \widehat{h}_4 - \widehat{L} + \widehat{h}_S
+ \widehat{h}_{yy}\right ) = 0 \ ,
\eeq

\noi $2aB^2 \times$ (A.5c) becomes

\bea
\label{4.13e}
&&[a\widehat{h}'_S ]' + a (N-1) {C' \over C} \left (
\widehat{h}'_4 + \widehat{h}'_S - \widehat{h}'_{yy} \right ) \nn \\
&&+ a {B^2
\over A^2} m^2 \widehat{h}_S + 2a {B^2 \over C^2} (N-2) \left ( 
\widehat{h}_S - (N-1) \widehat{h}_{yy}\right ) = 0 \ ,
\eea

\noi and $(-{A^2 \over m^2})\times$ (A.9) turns into

\beq \label{4.14e} \widehat{L}'  - \widehat{h}'_4 - \widehat{h}'_S
+\left ( {A' \over A} - {C' \over C}\right )  \widehat{h}_{s} + E'
\widehat{h}_{yy} = 0 \ . \eeq

\noi There is no need to reconsider eq. (A.5b) as well, since the five
equations (A.5b), (A.5c), (A.7)-(A.9) are related through the Bianchi
identity. The above equations (\ref{4.11e}-\ref{4.13e}) are coupled
wave equations for modes $\widehat{h}_4$, $\widehat{L}$, $\widehat{h}_S$
and $\widehat{h}_{yy}$. Out of the four independent equations
(\ref{4.11e})-(\ref{4.14e}) we can construct two more dependent
equations, that prove to be useful subsequently: Multiplying eq.
(\ref{4.14e}) by $a$, taking the derivative with respect to $y$, using
eqs. (\ref{4.11e})-(\ref{4.13e}) and relations between the warp factors
$A$, $B$ and $C$ given in eq. (2.6), one can derive another equation
involving only first order derivatives with respect to $y$ of the four
modes:

\bea \label{4.15e} &&E' \widehat{h}'_4 + \left ( E' + {A' \over A} -
{C' \over C}\right ) \widehat{h}'_S + (N-2) {B^2 \over C^2}
\widehat{h}_S \nn \\ &&- (N-1) (N-2) {B^2 \over C^2} \widehat{h}_{yy} +
{B^2 \over A^2} m^2 \left ( \widehat{h}_4 - \widehat{L} + \widehat{h}_S
\right ) = 0 \ . \eea

Next, from $(N-1) {C' \over C} \times$ (eq. (\ref{4.12e}) --
eq. (\ref{4.11e})) + $3 {A' \over A} \times$ eq. (\ref{4.13e}) and
eliminating $\widehat{h}_{yy}$ using (\ref{4.14e}) one derives

\bea \label{4.16e} &&(N-1) {C' \over C} [a (\widehat{L}' -
\widehat{h}'_4)]' + 3 {A' \over A} [a \widehat{h}'_S]' + {B^2 \over
A^2} m^2 \left [ (N-1){C' \over C} a (\widehat{L} - \widehat{h}_4) + 3
{A' \over A} a \widehat{h}_S\right ]\nn \\ &&+ 6(N-2) {A'B^2 \over
AC^2E'} \left [ (N-1)a (\widehat{L}' - \widehat{h}'_4 - \widehat{h}'_S)
+ (N+2) {A' \over A} a \widehat{h}_S \right ] = 0\eea

\noi which will be used below.\par

It is straightforward to check that the above bulk equations are
invariant under the gauge transformations (\ref{4.7e}). In order
to evade this gauge ambiguity we
now proceed as follows: First we  construct three independent gauge
invariant fields out of the four 4d scalars above. Then we study
their bulk equations and solve for the two modes with respect to the third.
Thus we show that the bulk equations can be reduced to a unique second order
differential equation in $y$ for one independent mode. Finally, we demonstrate
that the junction conditions for this mode imply both Dirichlet and
Neumann boundary conditions on both branes, hence there are no gauge
invariant fluctuations at all. \par

Three independent gauge invariant fields $H_1$, $H_2$ and $H_3$ can be
defined as follows:

\bminiG{4.17e} \label{4.17ae} H_1 = a \left [ (N-1) {C' \over C} (
\widehat{L} - \widehat{h}_4) + 3{A' \over A} \widehat{h}_S\right ]
\eeeq \beeq \label{4.17be} H_2 = a \left [ \widehat{L} ' - \widehat{h}'
_4 + \left ( E' + {A' \over A}\right ) (\widehat{L}  - \widehat{h} _4)
+ 3 {A' \over A}  \widehat{h}_{yy}\right ] \eeeq \beeq \label{4.17ce}
H_3 = a \left [ 3 {A' \over A} \widehat{h}' _4 + \left ( 4 {A' \over A}
\left ( E' + {A' \over A}\right ) + m^2 {B^2 \over A^2}\right )
(\widehat{h}_4 - \widehat{L}) - 12 {A'^2 \over A^2}   - \widehat{h}
_{yy}\right ] \  . \emini

\noi The first $y$ derivative of $H_1$ can be written as

\beq \label{4.18e} H'_1 = a \left [ (N-1) {C' \over C} ( \widehat{L}' -
\widehat{h}'_4) + 3{A' \over A} \widehat{h}'_S + (N-1) (N-2) {B^2 \over
C^2} (\widehat{L} - \widehat{h}_4)\right ] \ . \eeq

\noi Here, and in the following, we use relations between the
warp factors $A$, $B$ and $C$ derived from eqs. (2.6).
Now, the bulk equation (\ref{4.14e}) can be used to show that

\beq \label{4.19e} H'_1 = E' H_2 + \left ( {A' \over A} - {C' \over C}
\right ) H_1 \ . \eeq

\noi Next, from the bulk equation (\ref{4.15e}) one can derive

\beq \label{4.20e} \left [ {(N+2) \over (N-1)} \ {A' \over A} \left (
E' + {A' \over A}\right ) + m^2 {B^2 \over A^2} \right ] H_1 + 3 {A'
\over A} \left ( E' +  {A' \over A} - {C' \over C} \right ) H_2 - E'
H_3 = 0 \ . \eeq

\noi From eqs. (\ref{4.19e}) and (\ref{4.20e}) one can obtain $H_2$ and
$H_3$ in terms of $H_1$ and $H'_1$, which is left as the only
remaining independent degree of freedom. Finally, after some algebra
the bulk equation (\ref{4.16e}) becomes an equation that contains $H_1$
only:

\bea \label{4.21e} E'H''_1 &+ &\left ( E'D' - 2(N-1) (N-2) {B^2 \over
C^2} \right ) H'_1\nn \\ &+& \left ( 2 (N+2) (N-2) {B^2 \over C^2} \
{A' \over A} + {B^2 \over A^2} E' m^2 \right ) H_1 = 0.\eea

It remains to study the junction (boundary) conditions, subject to
which eq. (\ref{4.21e}) has to be solved.\par

First, in terms of the above four 4d scalar non-zero modes the junction
conditions (A.14) -- (A.15) can be written as (here we neglect
the effect of induced Einstein terms, that are discussed in the next
section, and which are included for completeness in the junction
conditions in the appendix)

\bminiG{4.22e} \label{4.22ae}
\left [ - (N-1) \widehat{h}'_4 - (N-2) \widehat{h}'_S + (N-1) \left (
E' + {A' \over A} - {C' \over C}\right )  \widehat{h}_{yy} \right ]_i
= 0\ ,   \eeeq
\beeq \label{4.22be}
\left [ - 3\widehat{h}'_4 - 4\widehat{h}'_S + 4E' \widehat{h}_{yy} 
\right ]_i = 0\ , \eeeq 
\beeq \label{4.22ce}
\left [ \widehat{L}' - \widehat{h}'_4 - \widehat{h}'_S + E' 
\widehat{h}_{yy} \right ]_i = 0 \ .
\emini
 
Using the bulk equation (\ref{4.14e}) in (\ref{4.22ce}), one derives

\beq
\label{4.23e}
[\widehat{h}_S]_i = 0 \ .
\eeq

\noi (Here one needs

$$\left [ {A' \over A} - {C' \over C}\right ]_i \not= 0 \ ,$$

\noi which follows from the junction conditions (2.7).)\par

Next, from ${A' \over A} \times$ eq. (\ref{4.22be}) + ${C' \over C}
\times$ eq. (\ref{4.22ae}) and the use of the bulk equation
(\ref{4.15e}) one derives similarly

\beq \label{4.24e} \left [ \widehat{L} - \widehat{h}_4 \right ]_i = 0 \
. \eeq

 From $(N-1) \times$ eq. (\ref{4.22be}) - $3 \times$ eq.
(\ref{4.22ae}) one finds

\beq \label{4.25e} \left [ \widehat{h}'_S - (N-1) {C' \over C}
\widehat{h}_{yy} \right ]_i = 0 \ , \eeq

\noi and from the use of (\ref{4.25e}) in (\ref{4.22ce})

\beq \label{4.26e} \left [ \widehat{L}' - \widehat{h} '_4 + 3 {A' \over
A} \widehat{h}_{yy} \right ]_i = 0 \ . \eeq
Finally, from eqs. (\ref{4.23e})-(\ref{4.26e}) in eqs. (\ref{4.17ae})
and (\ref{4.18e}) one derives easily

\beq \label{4.27e} [H_1]_i = [H'_1]_i = 0 \ . \eeq

The second order bulk equation (\ref{4.21e}) for the only independent
gauge invariant fluctuation $H_1$, subject to the boundary conditions
(\ref{4.27e}) at both branes, has only the trivial solution $H_1 = 0$,
and then $H_2 = H_3 = 0$ follows from eqs. (\ref{4.19e}) and
(\ref{4.20e}).\par

Hence there are no physical (gauge invariant) fluctuations at all,
notably none with a negative mass$^2$, that would indicate an
instability of the present configuration.
Furthermore a similar analysis to the above, 
for $\widehat{h}_4$, $\widehat{h}_{yy}$ and $\widehat{h}_S$ shows that
there are no gauge independent zero modes which simply confirms the fact
that there are no parameters left free once the junction conditions are 
imposed. 

A possible hint for the absence of physical fluctuations in the bulk
stems from the situation in 5d brane worlds [2, 3], which suffer from a
massless radion, but which has no Kaluza-Klein tower [4, 5]. Here the
junction conditions eliminate the massless modes in the bulk, hence it
is plausible that there are no massive Kaluza-Klein modes left at all.

\section{The effect of induced Einstein terms}

The presence of induced stress-energy tensors on the branes is 
motivated, amongst others, as an unavoidable quantum effect of matter on
the branes \cite{13r}. The same argument can be put forward in favour
of the presence of induced Einstein terms on the branes \cite{ig}. It
is interesting to study how the present setup would react to the
presence of these contributions to the actions on the branes, which is
what we will present in this chapter.

First, the action (\ref{2.2e}) gets now replaced by

\bea \label{5.1e}
S &=& \int d^4x\ d^{N-1}zdy \left \{ \sqrt{-g^{(D)}} \ {1 \over 2 
\kappa_D} \ R^{(D)} \right .\nn \\
&&\left . + \sum_{i=1,2} \sqrt{-g^{(D-1)}} \ \delta (y - y_i) \left [ 
\Lambda_i + {1 \over 2 \kappa_i}\
R^{(D-1)} + \Gamma_{i}^{(ind)} \right ] \right \} \ ,\eea

\noi where the Einstein terms on the branes get multiplied with a priori
arbitrary (unknown) coefficients $\kappa_i^{-1}$. The only effect of
these terms are modifications of the junction conditions of the
gravitational field, i.e. of the warp factors $A$, $B$, and $C$. Instead
of (2.7) the junction conditions read now \cite{17r}

\bminiG{5.2e}
\label{5.2ae}
{1 \over \kappa_D B_i} \left ( - 3 \left [ {A'' \over A}\right ]_i - 
(N-1) \left [ {C'' \over
C}\right ]_i \right ) = - {(N-1)(N-2) \over 2 \kappa_i C_i^2} + 
\Lambda_i + \Lambda_i^{(ind, 4)} \ ,
\eeeq
\beeq
  \label{5.2be}
{1 \over \kappa_D B_i} \left ( - 4 \left [ {A'' \over A}\right ]_i - 
(N-2) \left [ {C'' \over
C}\right ]_i \right ) = - {(N-3)(N-2) \over 2 \kappa_i C_i^2} + 
\Lambda_i + \Lambda_i^{(ind,S)}  
\emini

\noi leading to

\beq \label{5.3e}
\sqrt{f_i} (1 - w) + {1 \over \sqrt{f_i}} (1 + w) =
\sigma_i \kappa_D \left ( {N-2 \over \kappa_iC_i} + C_i \left ( 
\Lambda_i^{(ind,S)} -
\Lambda_i^{(ind,4)}\right ) \right ) \ , 
\eeq

\noi instead of eq. (2.9a), whereas eq. (2.9b) remains unchanged.

Note that the counting of free parameters in the configuration of the
gravitational field versus the number of equations to be satisfied
remains unchanged, hence the need to perform one fine tuning among the
parameters in the action remains valid. But, the presence of the
induced Einstein terms on the branes will affect the parameters
$\alpha$, $r_0$ and $|{\widehat y}|$. An analytic solution of eqs.
(5.3) and (2.9b) for $\alpha$, $r_0$ and $|{\widehat y}|$ is
unfortunately impossible, but it is instructive to consider the case
where the induced stress-energy tensors $\Lambda_i^{ind}$ are switched
off, and where the coefficients $\kappa_i^{-1}$ are relatively small:

Once the dimensionless ratio $\varepsilon_i$ satisfies

\beq \label{5.4ae}
\varepsilon_i \equiv \kappa_D^2 \left | {\Lambda_i \over \kappa_i} 
\right | \ll 1 
\eeq

\noi on each brane, one finds

\beq
\label{5.5e}
{|\widehat{y}| \over r_0} \sim max(\varepsilon_i)\ ,
\eeq

\noi i.e. one is back in the "close brane limit" as in the case of small
induced stress-energy tensors. More precisely, the expressions (2.20a)
and (2.20b) for $\alpha$ and $r_0$, respectively, remain valid, whereas
for the distance ${\widehat y}$ between the branes one obtains

\beq \label{5.6e}
|\widehat{y}| = - {\kappa_D \over 2} \left ( {1 \over \kappa_1} + {1 
\over \kappa_2}\right ) \left (
{w-1 \over w+1}\right )^{n_C-{1\over 2}} \ .
\eeq

Of utmost importance is the minus sign on the
right-hand side of eq. (\ref{5.6e}), which implies that

\beq
\label{5.7e}
{1 \over \kappa_1} + {1 \over \kappa_2} < 0
\eeq

\noi (in the present case with vanishing induced  stress-energy tensors
$\Lambda_i^{ind}$). If eq. (\ref{5.7e}) is not satisfied, the junction
conditions cannot be satisfied by the metric in the bulk, which was
assumed to be $x^{\mu}$-independent. Hence we would expect that in this
case either the gravitational field becomes time-dependent, or the
branes have to be (anti-)de-Sitter branes, or both. 

If we insist on a time independent metric and on Poincare-branes, at
least one of the coefficients $\kappa_i^{-1}$ of the Einstein terms has
to have the "wrong" sign. A priori this is not a disaster; what counts,
is the absence of ghosts (and tachyons) in the spectrum of gravitational
fluctuations. Here, however, we will encounter problems:

Now, in the presence of Einstein terms on the branes, the effective 4d
action for 4d gravity gets modified. It is still of the Einstein form
(\ref{3.10e}), but the expression (\ref{3.11e}) for the effective 4d
(inverse) Planck scale changes. Now it is of the form

\beq \label{5.8e}
{1 \over \kappa_4} = \Omega^{N-1} \left \{ {2 \over \kappa_D} 
\int_{0}^{\widehat{y}} dy\ A^2(y) \ B(y) \ C(y)^{N-1}
+ \sum_{i=1,2} {A_i^2 C_i^{N-1} \over \kappa_i} \right \}
\eeq

\noi with two additional terms due to the Einstein terms on the branes.
In the close brane limit all terms on the right hand side of eq.
(\ref{5.8e}) can be computed explicitely, using eqs. (2.20) for
$\alpha$, $r_0$ and (\ref{5.6e}) for $\widehat{y}$. The net result is
that once eq. (\ref{5.7e}) is satisfied, eq. (\ref{5.8e}) gives always
a negative gravitational coupling, i.e. a graviton ghost, since the
negative contribution to the second term on the right hand side is
never completely cancelled by the positive integral $dy$ across the
bulk. Consequently Einstein terms on the branes only, without induced
stress-energy tensors on the branes, do not allow for a consistent
configuration. Only in the presence of induced stress-energy tensors,
Einstein terms on the branes would be allowed, provided that their
coefficients $\kappa_i^{-1}$ are not too large.

We have also checked that the absence of gauge invariant scalar
fluctuations in the bulk, as discussed in section 4, remains valid in
the presence of such Einstein terms on the branes.

\section{Conclusions and Outlook}

The essential result of this paper is a new  stabilization mechanism
both for the radion and the overall modulus (the size of the
compactified extra dimensions parallel to the branes) in higher
dimensional brane worlds. It is based on the additional components of
the junction conditions, that arise once extra dimensions parallel to
the branes are compactified, and notably on the presence of induced
terms on the branes. \par

We have studied this mechanism in detail in the simplest possible 
setup: Higher dimensional brane worlds with an empty bulk, $N-1$ extra
dimensions ($N>2$) parallel to the  branes compactified on $S^{(N-1)}$,
hence a resulting geometry of the form $R^{(3,1)} \times S^{(N-1)} 
\times S^1/Z_2$. \par

Matter fields play only an indirect role, namely through the terms on
the branes (that affect the junction conditions) induced by their
quantum fluctuations. Among the possible induced terms on the branes
we have confined ourselves to constant (diagonal) stress-energy tensors
and Einstein scalars. We have seen that Einstein scalars cannot play a
dominant role; if their coefficients (the sum of which must be
negative, cf. eq. (2.19)) are too large, they turn the graviton into a
ghost. The induced stress-energy tensors resemble to induced
cosmological constants on the branes with the important difference,
however, that the diagonal elements along $R^{(3,1)}$ and $S^{(N-1)}$ 
are generically different. Geometrically this goes hand in hand with
the fact that each subspace is of different curvature. This difference
turned out to be crucial for the configuration to be consistent, i.e.
to satisfy the junction conditions for a real non-singular
time-independent metric. \par

We have attributed the dynamical origin of the difference among  the
diagonal elements of the stress-energy on the branes to  matter-induced
radiative corrections as studied in ref. \cite{13r}. Its precise
dependence on the matter spectrum on the branes and the 
compactification radius needs eventually additional investigations.
Such a difference among  the diagonal elements of the stress-energy on
the branes could actually be generated at the  classical level as well,
if form fields with the appropriate rank would live on the branes.\par

An additional necessary condition for the stability of the 
configuration is the positivity of all masses$^2$ of all fluctuations
associated to deformations of the configuration. Our somewhat
surprising result is that in the present setup there are no such
fluctuations (maintaining the shape of $S^{(N-1)}$, but corresponding
to variations of the warp factors across the bulk) at all. All scalar
fluctuations are frozen.

One should remark that the coupled Einstein field
equations in the bulk and the junction conditions on the branes are
effects of self-gravity "sourced" by the stress-energy tensors on the
branes, and that it is thus pure Einstein gravity that yields the
rigidity of the whole setup (at least for scalar fluctuations). \par

At present we have assumed an empty bulk. The effects of 
additional fields in the bulk, and generalizations of the compact
manifold $M^{N-1}$ have to be studied; only then the present mechanism
can eventually be applied to brane worlds motivated by string/M theory.

\section*{Acknowledgements}

We are particularly grateful to J. Mourad and V. Rubakov for helpful
discussions and comments. C.C. thanks J.-F. Dufaux for discussions on
radion related issues.

\section*{Appendix A}

In this appendix we give the components  $R_{\mu\nu}$, $R_{yy}$,
$\gamma^{\alpha\beta}R_{\alpha\beta}$ and $R_{\mu  y}$ of the Ricci
tensor linear in the fluctuations $\widehat{h}_{\mu\nu}$,
$\widehat{h}_{yy}$ and  $\widehat{h}_S$ as described in eqs.
(\ref{3.4e}) -- (\ref{3.6e}). For simplicity we consider
only zero  modes on $S^{(N-1)}$. Subsequently we give the junction
conditions for the fluctuations as derived from eqs. (\ref{2.3e})
linear in the fluctuations. \par

We use the following notations: the 4d Laplacian is denoted by 
$\sq^{(4)}$, with

$$\sq^{(4)} = g^{\mu\nu} \partial_{\mu}\partial_{\nu} = A^{-2}(y)
\eta^{\mu\nu} \partial_{\mu}\partial_{\nu} \ . \eqno({\rm A.}1)$$

Derivatives with respect to $y$ are denoted by primes. $D'$ and $E'$
denote the following combinations of $y$ derivatives of the warp
functions $A(y)$, $B(y)$ and $C(y)$ (as given in eqs. (\ref{2.6e})):

$$D' = 4 {A' \over A} + (N-1) {C' \over C} - {B' \over B} \ ,
\qquad E' = 3 {A' \over A} + (N-1) {C' \over C} \ .
\eqno({\rm A.}2)$$

\noi The 4d trace of $\widehat{h}_{\mu\nu}$ is denoted by
$\widehat{h}_4$:

$$\widehat{h}_4 = \eta^{\mu\nu} \ \widehat{h}_{\mu\nu} \left ( = 
g^{\mu\nu} h_{\mu\nu}\right ) \ . \eqno({\rm A.}3)$$

\noi Finally, $K_{\mu}$ is defined as (with $\partial^{\nu} = A^{-2} 
\eta^{\nu\mu} \partial_{\mu}$)

$$K_{\mu} = \partial^{\nu} \widehat{h}_{\mu\nu} - {1 \over 2A^2} 
\partial_{\mu} \left ( \widehat{h}_4
+ \widehat{h}_S + \widehat{h}_{yy} \right ) \ . \eqno({\rm A.}4)$$

\noi Then we have
\bea
A^{-2} R_{\mu\nu} &=& {1 \over 2} \sq^{(4)} \widehat{h}_{\mu\nu} - {1 
\over 2} \left (
\partial_{\mu} K_{\nu} +  \partial_{\nu} K_{\mu} \right )\nn \\
&& + {1 \over 2B^2} \left [ \widehat{h}_{\mu\nu}'' + 
\widehat{h}_{\mu\nu}'D' + \eta_{\mu\nu} {A'
\over A} \left ( \widehat{h}_4' + \widehat{h}_S' - \widehat{h}_{yy}' 
\right ) \right ] \ , \hskip 3,1cm ({\rm A.5a}) \nn \\
R_{yy} &=& B^2 \sq^{(4)} \widehat{h}_{yy} - \widehat{h}_{yy}' \left 
(D' + {B' \over B} \right ) +
\widehat{h}_4'' + \widehat{h}_4' \left ( 2 {A' \over A} - {B' \over 
B} \right ) + \widehat{h}_S'' \nn\\
&& +\widehat{h}_S' \left ( 2 {C' \over C} - {B' \over B}\right ) 
\ ,\hskip 8,1cm ({\rm A.5b}) \nn\\
C^{-2} \gamma^{\alpha\beta} R_{\alpha\beta} &=&  {1 \over 2}  \sq^{(4)}
\widehat{h}_S + {N-2 \over C^2} \widehat{h}_S - {(N-1)(N-2) \over C^2}
\widehat{h}_{yy}
\nn\\&&
+ {1 \over 2B^2} \left [ \widehat{h}_S'' + \widehat{h}_S' D' +  (N-1)
{C' \over C} \left ( \widehat{h}_4' + \widehat{h}_S' -
\widehat{h}_{yy}' \right ) \right ] 
\ ,\hskip 2,7cm ({\rm A.5c}) \nn\\
2R_{\mu y} &=& - A^2 \partial^{\nu} \widehat{h}_{\mu\nu}' + 
\partial_{\mu} \left ( \widehat{h}_4' + \widehat{h}_S' \right ) + \left
( {C' \over C} - {A' \over A}\right  ) \partial_{\mu} \widehat{h}_S
- E' \partial_{\mu} \widehat{h}_{yy}\ .  \hskip 1,6cm ({\rm A.5d}) \nn
\eea

Actually we only need the components of the Ricci tensor contracted 
over $\mu$ and $\nu$. Then,
instead of the vector $K_{\mu}$ in (A.4), only the combination

$$L = A^2 \ \partial^{\mu} \partial^{\nu} \widehat{h}_{\mu\nu} 
\eqno({\rm A.}6)$$

\noi appears:
\bea
g^{\mu\nu} R_{\mu\nu}&=& A^{-2} \eta^{\mu\nu} R_{\mu\nu} = \sq^{(4)}
\widehat{h}_4 - L + {1 \over 2} \sq^{(4)} \left ( \widehat{h}_S +
\widehat{h}_{yy} \right )\nn \\
&&+ {1 \over 2B^2} \left [ \widehat{h}_4'' + \widehat{h}_4' \left ( 
D' + 4 {A' \over
A}\right ) + 4 {A' \over A} \left ( \widehat{h}_S' - 
\widehat{h}_{yy}'\right ) \right ]\ ,\hskip 4,0cm ({\rm A.7}) \nn 
\eea
\bea
\partial^{\mu} \partial^{\nu} R_{\mu\nu}&=& {1 \over 2} \sq^{(4)} 
\left [ \sq^{(4)} \ \widehat{h}_4 - L + \sq^{(4)} \left ( \widehat{h}_S
+ \widehat{h}_{yy} \right ) \right ]\nn \\&&
+ {1 \over 2B^2} \left [ L'' + L'\left (D' + 4 {A' \over
A}\right  ) + 4 {A'^2 \over A} L 
 + {A' \over A} \sq^{(4)} \left ( \widehat{h}_4' + 
\widehat{h}_S' - \widehat{h}_{yy}'\right ) \right ] \ ,\hskip 0,8cm  
({\rm A.}8) \nn
\eea
$$2\partial^{\mu}R_{\mu y} = - L' - 2 {A' \over A} L + \sq^{(4)} 
\left ( \widehat{h}_4' + \widehat{h}_S'\right )
+ \left ( {C' \over C} - {A' \over A}\right ) \sq^{(4)}  
\widehat{h}_S - \left( 3 {A' \over A}+(N-1){C' \over C}\right) 
\sq^{(4)}  \widehat{h}_{yy} .\eqno({\rm A.}9)$$

Next we turn to the junction conditions for the fluctuations. For
completeness we include here the Einstein terms on the branes. Then the
Einstein equations, including the brane terms from the action
(\ref{5.1e}), read (instead of (\ref{2.3e}))

$${1 \over \kappa_D} G_{\mu\nu}^{(D)} = \sum_{i=1,2} \sqrt{{g^{(D-1)} 
\over g^{(D)}}} \delta (y-y_i)
\left [ g_{\mu\nu} \Lambda_i - {1 \over \kappa_i} G_{\mu\nu}^{(D-1)} 
+ g_{\mu\nu} \Lambda_i^{(ind,4)}
\right ] \ ,  \eqno({\rm A.10a})$$
$${1 \over \kappa_D} G_{\alpha\beta}^{(D)} = \sum_{i=1,2} 
\sqrt{{g^{(D-1)} \over g^{(D)}}} \delta (y-y_i)
\left [ g_{\alpha\beta} \Lambda_i - {1 \over \kappa_i} 
G_{\alpha\beta}^{(D-1)} + g_{\alpha\beta}
\Lambda_i^{(ind,S)} \right ] \ .\eqno({\rm A.10b})$$

The induced Einstein tensor $G_{AB}^{(D-1)}$ ($A \not = y$ and $B \not=
y$) is  constructed from $g_{AB}$ with $A \not= y$ and $B \not= y$. 
Now also the right hand sides of eqs. (A.10) have to be expanded to
linear order in the fluctuations. (As stated in chapter 4, brane
bending modes $\beta_i(x^\mu)$ are omitted using a corresponding
gauge.) There are terms on the left-hand side of eq. (A.10), to linear
order in the fluctuations, which are proportional to the jumps in the
$y$  derivatives of the background metric. Nearly all of these terms
cancel on both sides once the  junction conditions (\ref{5.2e}) for the
background metric are used; only terms proportional to 
$\widehat{h}_{yy}$ are left on the left-hand side. \par

Again it is useful to use the quantities $D'$, $E'$ defined in eq.
(A.2). Instead of the vector $K_{\mu}$ in eq. (A.4), however, it is
more useful to introduce $\widehat{K}_{\mu}$  without
$\widehat{h}_{yy}$ (which does not enter the right-hand side of the
junction conditions):

$$\widehat{K}_{\mu} = \partial^{\nu} \ \widehat{h}_{\nu \mu} - {1
\over  2A^2} \ \partial_{\mu} \left ( \widehat{h}_4 + \widehat{h}_S
\right ) \ . \eqno({\rm A.}11)$$

After using the junction conditions (\ref{5.2ae}) for the background 
metric, as stated above, the
$(\mu, \nu)$ components of the junction conditions for the 
fluctuations read after multiplication
with $B_i/A_i^2$ (where the index $i$ denotes the argument $y = y_i)$:

$${\sigma_i \over B_i \kappa_D} \left [ \widehat{h}_{\mu \nu}' + 
\eta_{\mu\nu} \left (E' \widehat{h}_{yy} -  \widehat{h}_{4}' -
\widehat{h}_S'\right ) \right ]_i$$
$$ = {1 \over 2 \kappa_i} \Big [- \sq^{(4)} \ \widehat{h}_{\mu \nu} +
\left ( \partial_{\mu} \widehat{K}_{\nu} + \partial_{\nu}
\widehat{K}_{\mu} \right )
+ \eta_{\mu\nu} \left ( \sq^{(4)} \ \widehat{h}_4 + \sq^{(4)} \ 
\widehat{h}_S - L + {N-2 \over C^2} \widehat{h}_S\right ) \Big
]_i \eqno({\rm A.}12)$$

\noi where $\sigma_1 = + 1$, $\sigma_2 = - 1$. \par

Since we are only interested in the mode $\widehat{h}_S =
\gamma^{\alpha\beta} \widehat{h}_{\alpha\beta}$ with indices in
$S^{(N-1)}$, it is  sufficient to give the $(\alpha , \beta )$
components contracted with $\gamma^{\alpha\beta}$. After using the 
junction conditions (\ref{5.2be}), and after multiplication with
$B_i/C_i^2$, one obtains

$${\sigma_i \over B_i \kappa_D} \left [ - (N-1) \widehat{h}_4' -  (N-2)
\widehat{h}_S' + (N-1) \widehat{h}_{yy} \left ( 4{A' \over A} + (N-2)
{C' \over C} \right ) \right ]_i$$
$$= {1 \over 2 \kappa_i} \left [ (N-1) \left ( \sq^{(4)}  \widehat{h}_4
- L \right ) + (N-2) \sq^{(4)}  \widehat{h}_S + {(N-2) (N-3) \over
2C^2} \widehat{h}_S \right ]_i \ . \eqno({\rm A.}13)$$

It is also useful to have the expressions for contracted $(\mu ,  \nu)$
components of the junction conditions (A.12).
For the contraction of (A.12) with $\eta^{\mu\nu}$ one finds

$$\sigma_i \left [ - 3  \widehat{h}_4' - 4\widehat{h}_S' + 4 E'
\widehat{h}_{yy} \right ]_i
= {B_i\kappa_D \over \kappa_i} \left [ \sq^{(4)} \widehat{h}_4 - L 
+ {3 \over 2} \sq^{(4)}
\widehat{h}_S + {2(N-2) \over C^2} \widehat{h}_S \right ]_i \ 
,\eqno({\rm A.}14)$$

\noi and from its contraction with derivatives $A^2 \partial^{\mu}
\partial^{\nu}$,
$$\sigma_i \left [ L' + 2 {A' \over A} L + \sq^{(4)} \left ( E'
\widehat{h}_{yy} - \widehat{h}_4' - \widehat{h}_S'  \right ) \right ] =
{B_i \kappa_D \over \kappa_i} \left [ {(N-2) \over 2C^2} \sq^{(4)} 
\widehat{h}_S \right ]_i .$$
$$\eqno({\rm A.}15)$$

In the case of massive modes as discussed in chapter 4 it is useful to
remember that, on plane waves of the form (\ref{4.4e}), the action of
$\sq^{(4)}$ becomes
$$\sq^{(4)} = - A^{-2}(y)\ \eta^{\mu\nu} k_{\mu}k_{\nu} = 
A^{-2}(y)\ m^2 \ . \eqno({\rm A.}16)$$


\begin{thebibliography}{99}

\bibitem{early}
V.~A.~Rubakov and M.~E.~Shaposhnikov,
``Do We Live Inside A Domain Wall?,''
Phys.\ Lett.\ B {\bf 125}, 136 (1983).

K.~Akama,
``An Early Proposal Of 'Brane World',''
Lect.\ Notes Phys.\  {\bf 176}, 267 (1982)
[arXiv:hep-th/0001113].

M.~Visser,
``An Exotic Class Of Kaluza-Klein Models,''
Phys.\ Lett.\ B {\bf 159}, 22 (1985)
[arXiv:hep-th/9910093].

E.~J.~Squires,
``Dimensional Reduction Caused By A Cosmological Constant,''
Phys.\ Lett.\ B {\bf 167}, 286 (1986).

\bibitem{1r}
L.~Randall and R.~Sundrum,
``A large mass hierarchy from a small extra dimension,''
Phys.\ Rev.\ Lett.\  {\bf 83} (1999) 3370
[arXiv:hep-ph/9905221].

\bibitem{2r}
P.~Binetruy, C.~Deffayet and D.~Langlois,
``Non-conventional cosmology from a brane-universe,''
Nucl.\ Phys.\ B {\bf 565}, 269 (2000)
[arXiv:hep-th/9905012].

\bibitem{3r}
J.~Garriga and T.~Tanaka,
``Gravity in the brane-world,''
Phys.\ Rev.\ Lett.\  {\bf 84}, 2778 (2000)
[arXiv:hep-th/9911055].

\bibitem{4r}
C.~Charmousis, R.~Gregory and V.~A.~Rubakov,
``Wave function of the radion in a brane world,''
Phys.\ Rev.\ D {\bf 62}, 067505 (2000)
[arXiv:hep-th/9912160].

\bibitem{5r}
P.~Horava and E.~Witten,
``Heterotic and type I string dynamics from eleven dimensions,''
Nucl.\ Phys.\ B {\bf 460}, 506 (1996)
[arXiv:hep-th/9510209].

\bibitem{6r}
P.~Horava and E.~Witten,
``Eleven-Dimensional Supergravity on a Manifold with Boundary,''
Nucl.\ Phys.\ B {\bf 475}, 94 (1996)
[arXiv:hep-th/9603142].

\bibitem{7r}
A.~Lukas, B.~A.~Ovrut, K.~S.~Stelle and D.~Waldram,
``Heterotic M-theory in five dimensions,''
Nucl.\ Phys.\ B {\bf 552}, 246 (1999)
[arXiv:hep-th/9806051].

A.~Lukas, B.~A.~Ovrut, K.~S.~Stelle and D.~Waldram,
``The universe as a domain wall,''
Phys.\ Rev.\ D {\bf 59}, 086001 (1999)
[arXiv:hep-th/9803235].

\bibitem{8r}
P.~Candelas and S.~Weinberg,
``Calculation Of Gauge Couplings And Compact Circumferences From
Selfconsistent Dimensional Reduction,'' Nucl.\ Phys.\ B {\bf 237}, 397
(1984).\\

Y.~I.~Kogan and N.~A.~Voronov,
``Spontaneous Compactification in the Kaluza-Klein Models and the
Casimir Effect,''
JETP Lett.\  {\bf 38}, 311 (1983).

\bibitem{9r}
W.~D.~Goldberger and M.~B.~Wise,
``Modulus stabilization with bulk fields,''
Phys.\ Rev.\ Lett.\  {\bf 83}, 4922 (1999)
[arXiv:hep-ph/9907447].

\bibitem{10r}
P.~Kanti, I.~I.~Kogan, K.~A.~Olive and M.~Pospelov,
``Cosmological 3-brane solutions,''
Phys.\ Lett.\ B {\bf 468}, 31 (1999)
[arXiv:hep-ph/9909481].

P.~Kanti, I.~I.~Kogan, K.~A.~Olive and M.~Pospelov,
``Single-brane cosmological solutions with a stable compact extra 
dimension,''
Phys.\ Rev.\ D {\bf 61}, 106004 (2000)
[arXiv:hep-ph/9912266].

\bibitem{11r}
O.~DeWolfe, D.~Z.~Freedman, S.~S.~Gubser and A.~Karch,
``Modeling the fifth dimension with scalars and gravity,''
Phys.\ Rev.\ D {\bf 62}, 046008 (2000)
[arXiv:hep-th/9909134].

\bibitem{12r}
J.~Garriga, O.~Pujolas and T.~Tanaka,
``Radion effective potential in the brane-world,''
Nucl.\ Phys.\ B {\bf 605}, 192 (2001)
[arXiv:hep-th/0004109].

W.~D.~Goldberger and I.~Z.~Rothstein,
``Quantum stabilization of compactified AdS(5),''
Phys.\ Lett.\ B {\bf 491}, 339 (2000)
[arXiv:hep-th/0007065].


S.~Mukohyama,
``Quantum effects, brane tension and large hierarchy in the brane
world,''
Phys.\ Rev.\ D {\bf 63}, 044008 (2001)
[arXiv:hep-th/0007239].

R.~Hofmann, P.~Kanti and M.~Pospelov,
``(De-)stabilization of an extra dimension due to a Casimir force,''
Phys.\ Rev.\ D {\bf 63}, 124020 (2001)
[arXiv:hep-ph/0012213].

A.~Flachi and D.~J.~Toms,
``Quantized bulk scalar fields in the Randall-Sundrum brane-model,''
Nucl.\ Phys.\ B {\bf 610}, 144 (2001)
[arXiv:hep-th/0103077].

%\cite{Saharian:2002bw}
A.~A.~Saharian and M.~R.~Setare,
 ``The Casimir effect on background of conformally flat brane-world
geometries,
Phys.\ Lett.\ B {\bf 552}, 119 (2003)
[arXiv:hep-th/0207138].
%%CITATION = HEP-TH 0207138;%%

\bibitem{13r} 
Z.~Chacko and A.~E.~Nelson,
``A solution to the hierarchy problem with an infinitely large extra 
dimension and moduli stabilization,''
Phys.\ Rev.\ D {\bf 62} (2000) 085006
[arXiv:hep-th/9912186].

\bibitem{fl} 
A.~Flachi, J.~Garriga, O.~Pujolas and T.~Tanaka,
``Moduli stabilization in higher dimensional brane models,''
JHEP {\bf 0308}, 053 (2003)
[arXiv:hep-th/0302017].\\
J.~Martin, G.~N.~Felder, A.~V.~Frolov, M.~Peloso and L.~Kofman,
``Braneworld dynamics with the BraneCode''
arXiv:hep-th/0309001.\\
J.~Lesgourgues and L.~Sorbo, ``Goldberger-Wise variations: Stabilizing
brane models with a bulk scalar'' arXiv:hep-th/0310007.

\bibitem{14r}
H.~Collins and B.~Holdom,
``Linearized gravity about a brane,''
Phys.\ Rev.\ D {\bf 62}, 124008 (2000)
[arXiv:hep-th/0006158].

\bibitem{15r}
G.~R.~Dvali, G.~Gabadadze and M.~Porrati,
``4D gravity on a brane in 5D Minkowski space,''
Phys.\ Lett.\ B {\bf 485}, 208 (2000)
[arXiv:hep-th/0005016].

\bibitem{ig}
S.~L.~Adler, ``Order R Vacuum Action Functional In Scalar Free Unified
Theories With Spontaneous Scale Breaking,'' Phys.\ Rev.\ Lett.\  {\bf
44}, 1567 (1980).\\
S.~L.~Adler,
``A Formula For The Induced Gravitational Constant,''
Phys.\ Lett.\ B {\bf 95}, 241 (1980).\\
S.~L.~Adler, ``Einstein Gravity As A Symmetry Breaking Effect In
Quantum Field Theory,'' Rev.\ Mod.\ Phys.\  {\bf 54}, 729 (1982)
[Erratum-ibid.\  {\bf 55}, 837 (1983)].\\
A.~Zee, ``Calculating Newton's Gravitational Constant In Infrared
Stable Yang-Mills Theories,'' Phys.\ Rev.\ Lett.\  {\bf 48}, 295
(1982).\\
N.~N.~Khuri,
``An Upper Bound For Induced Gravitation,''
Phys.\ Rev.\ Lett.\  {\bf 49}, 513 (1982).\\
N.~N.~Khuri,
``The Sign Of The Induced Gravitational Constant,''
Phys.\ Rev.\ D {\bf 26}, 2664 (1982).\\

\bibitem{Deffayet}
C.~Deffayet,
``Cosmology on a brane in Minkowski bulk,''
Phys.\ Lett.\ B {\bf 502}, 199 (2001)
[arXiv:hep-th/0010186].


\bibitem{Luty}
M.~A.~Luty, M.~Porrati and R.~Rattazzi,
``Strong interactions and stability in the DGP model,''
JHEP {\bf 0309}, 029 (2003)
[arXiv:hep-th/0303116].\\
V.~A.~Rubakov,
``Strong coupling in brane-induced gravity in five dimensions,''
arXiv:hep-th/0303125.

\bibitem{sc}
S.~L.~Dubovsky and M.~V.~Libanov,
``On brane-induced gravity in warped backgrounds,''
JHEP {\bf 0311}, 038 (2003)
[arXiv:hep-th/0309131].\\
M.~Porrati and J.~W.~Rombouts,
``Strong Coupling vs. 4-D Locality in Induced Gravity,''
arXiv:hep-th/0401211.

\bibitem{16r}
R.~Gregory,
``Cosmic p-Branes,''
Nucl.\ Phys.\ B {\bf 467}, 159 (1996)
[arXiv:hep-th/9510202].

\bibitem{roberto}
C.~Charmousis, R.~Emparan and R.~Gregory,
``Self-gravity of brane worlds: A new hierarchy twist,''
JHEP {\bf 0105}, 026 (2001)
[arXiv:hep-th/0101198].

\bibitem{17r}
U.~Ellwanger,
``Blown-up p-branes and the cosmological constant,''
JCAP {\bf 0311}, 013 (2003) 
[arXiv:hep-th/0304057].

\end{thebibliography}
\end{document}